\documentclass{epl}

\title{Confining Strings in the\\ 
Abelian-Projected SU(3)-Gluodynamics}
\author{D. Antonov\inst{1,2}}
\institute{
  \inst{1} INFN-Sezione di Pisa, Universit\'{a} degli studi di Pisa,
 Dipartimento di Fisica, Via Buonarroti, 
 2 - Ed. B - 56127 Pisa, Italy\\
  \inst{2} Institute of Theoretical 
 and Experimental Physics, B. Cheremushkinskaya 25, 
 117 218 Moscow, Russia
}

\begin{document}

\maketitle

\begin{abstract}
String representation of the Wilson loop in 
3D Abelian-projected $SU(3)$-gluodynamics is constructed 
in the approximation that 
Abelian-projected monopoles form a gas. Such an assumption is 
much weaker 
than the standard one, demanding the monopole condensation.
It is demonstrated that the summation over world sheets,
bounded by the contour of the Wilson loop, is realized 
by the summation over branches of a certain effective 
multivalued potential of the monopole densities.
Finally, by virtue of the so-constructed representation of the Wilson
loop in terms of the monopole densities, this quantity is 
evaluated in the approximation of a dilute monopole gas, which 
makes confinement in the model under study manifest. 
\end{abstract}

On the way of constructing the string representation 
of $SU(3)$-gluodynamics by means of the method 
of Abelian projections~\cite{abpr} (see Ref.~\cite{rev} for 
recent reviews), the 
main results have up to now been obtained under the assumption 
of the monopole condensation. Indeed, this assumption leads 
to an effective Ginzburg-Landau type theory~\cite{suzuki}, whose 
string representation can further be investigated~\cite{progress}
analogously to that of the usual dual superconductor~\cite{thmand}. 
On the other hand, recently string representation of the 
Abelian-projected $SU(2)$-gluodynamics has been derived~\cite{JPG} 
under a weaker assumption, which states that Abelian-projected 
monopoles form a gas, rather than condense into the dual Higgs field. 
Such a way of treating Abelian-projected monopoles in the 
$SU(2)$-gluodynamics 
makes the 
string representation of the Wilson loop in this theory (which describes 
an external test particle, electrically charged {\it w.r.t.} the 
$U(1)$ Cartan subgroup of $SU(2)$) similar to that of 
the Wilson loop in compact QED~\cite{confstr}.
The aim of the present 
Letter is to extend the results of Ref.~\cite{JPG} 
to the case of Abelian-projected 
$SU(3)$-gluodynamics in 2+1 dimensions and finally to emphasize
confinement in this theory in the sense of the Wilson area law~\cite{wil}.

Let us start our analysis with considering  
the pure monopole contribution to the action of this theory,  
keeping for a while aside the noncompact part of diagonal 
fields. (The off-diagonal fields are as usual 
disregarded on the basis of the so-called Abelian dominance 
hypothesis~\cite{abdom}. That is because they are argued to become 
very massive (and thus short-ranged) and therefore irrelevant 
to the IR region, where confinement holds.) 
The partition function describing 
the grand canonical ensemble of monopoles has the form~\cite{2}

\begin{equation}
\label{1}
{\cal Z}=1+\sum\limits_{N=1}^{\infty}
\frac{\zeta^N}{N!}\left(\prod\limits_{a=1}^{N}
\int d^3z_a\sum\limits_{\alpha_a=\pm 1,\pm 2,\pm 3}^{}\right)
\exp\left[-\frac{g_m^2}{4\pi}\sum\limits_{a<b}^{}
\frac{\vec q_{\alpha_a}\vec q_{\alpha_b}}{\left|\vec z_a-
\vec z_b\right|}\right].
\end{equation}
Here, $g_m$ is the magnetic coupling constant, related to the 
QCD coupling constant $g$ according to the equation $gg_m=4\pi$, 
$\zeta\propto\exp\left(-{\rm const.}/g^2\right)$ 
is the fugacity (Boltzmann factor) of a single monopole, and 
$\vec q_{\alpha_a}$'s are the nonzero weights of the zero 
triality adjoint representation of ${}^{*}SU(3)$. These weights are 
defined as 
$\vec q_1=(1/2,\sqrt{3}/2)$,
$\vec q_2=(-1,0)$,
$\vec q_3=(1/2,-\sqrt{3}/2)$,
$\vec q_{-\alpha}=-\vec q_\alpha$.
It is worth noting that for $\vec\lambda=(\lambda_3,
\lambda_8)$, where in the Gell-Mann basis $\lambda_3={\,}{\rm diag}{\,}
(1,-1,0)$, $\lambda_8={\,}{\rm diag}{\,}\left(1/\sqrt{3},1/\sqrt{3}, 
-2/\sqrt{3}\right)$, 
the following relations hold: $\vec q_1\vec\lambda={\,}
{\rm diag}{\,}(1,0,-1)$, $\vec q_2\vec\lambda={\,}{\rm diag}{\,}
(-1,1,0)$, $\vec q_3\vec\lambda={\,}{\rm diag}{\,}(0,-1,1)$.
Therefore for every $\alpha=\pm 1,\pm 2,\pm 3$, one has 
$\vec q_\alpha\vec\lambda=\hat n$, 
where $\hat n$ is a certain traceless diagonal matrix with 
the elements $0,\pm 1$. This matrix can thus be written as 
$\hat n=w\lambda_3w^{-1}$, where $w$ is any of the six elements 
of the permutation group $S_3$. 

Equation~(\ref{1}) can be represented as 

\begin{equation}
\label{aux}
{\cal Z}=1+\sum\limits_{N=1}^{\infty}
\frac{\zeta^N}{N!}\left(\prod\limits_{a=1}^{N}
\int d^3z_a\sum\limits_{\alpha_a=\pm 1,\pm 2,\pm 3}^{}\right)
\exp\left[-\frac{2\pi}{g^2}\int d^3xd^3y\vec\rho_{\rm gas}(\vec x)
\frac{1}{|\vec x-\vec y{\,}|}\vec\rho_{\rm gas}(\vec y)\right],
\end{equation}
or further

$$
{\cal Z}=\int {\cal D}\vec\chi\exp\left[-\frac12\int d^3x\left(\nabla
\vec\chi\right)^2\right]\times$$

\begin{equation}
\label{2}
\times\left[1+\sum\limits_{N=1}^{\infty}
\frac{\zeta^N}{N!}\left(\prod\limits_{a=1}^{N}
\int d^3z_a\sum\limits_{\alpha_a=\pm 1,\pm 2,\pm 3}^{}
\right)
\exp\left(ig_m
\int d^3x\vec\chi\vec\rho_{\rm gas}\right)\right].
\end{equation}
Here, $\vec\rho_{\rm gas}(\vec x)=\sum\limits_{a=1}^{N}\vec q_{\alpha_a}
\delta\left(\vec x-\vec z_a\right)$ stands for the  
density of the monopole gas, and the measure ${\cal D}\vec\chi$
is normalized by the condition 

$$\int {\cal D}\vec\chi\exp\left[
-\frac12\int d^3x\left(\nabla\vec\chi\right)^2\right]=1.$$
Equation~(\ref{2}) thus yields the following representation 
for the partition function~(\ref{1}):

\begin{equation}
\label{3}
{\cal Z}=\int {\cal D}\vec\chi\exp\left\{-\int d^3x\left[
\frac12\left(\nabla\vec\chi\right)^2-
\zeta\sum\limits_{\alpha=\pm 1,\pm 2,\pm 3}^{}
\exp\left(ig_m\vec q_\alpha\vec\chi
\right)\right]\right\},
\end{equation}
or, in the form analogous to the one of compact QED,  

\begin{equation}
\label{new}
{\cal Z}=\int {\cal D}\vec\chi\exp\left\{-\int d^3x\left[
\frac12\left(\nabla\vec\chi\right)^2-
2\zeta\sum\limits_{\alpha=1}^{3}\cos\left(g_m\vec q_\alpha\vec\chi
\right)\right]\right\}.
\end{equation} 
Denoting $\vec q_\alpha\vec\chi$, $\alpha=1,2,3$, 
by $\chi_\alpha$ and performing 
the rescaling $\chi_\alpha^{\rm new}=\sqrt{\frac23}
\chi_\alpha^{\rm old}$, we can represent the partition 
function~(\ref{new}) as 

\begin{equation}
\label{very}
{\cal Z}=\int\left(\prod\limits_{\alpha=1}^{3} {\cal D}\chi_\alpha
\right)\delta\left(\sum\limits_{\alpha=1}^{3}\chi_\alpha\right)
\exp\left\{-\int d^3x
\left[\frac12\left(\nabla\chi_\alpha
\right)^2
-2\zeta\sum\limits_{\alpha=1}^{3}\cos\left(g_m\sqrt{\frac32}
\chi_\alpha\right)\right]\right\}.
\end{equation}
Integrating out one of the fields $\chi_\alpha$'s, {\it e.g.}
for concreteness $\chi_3$, and denoting 
$\xi_1=\sqrt{\frac32}(\chi_1+\chi_2)$, $\xi_2=
\frac{1}{\sqrt{2}}(\chi_1-\chi_2)$, we get for the 
partition function~(\ref{very}) the following expression: 

$$
{\cal Z}=\int {\cal D}\xi_1 {\cal D}\xi_2\exp\left\{-\int d^3x\left[
\frac12(\nabla\xi_1)^2+\frac12(\nabla\xi_2)^2-\right.\right.$$

\begin{equation}
\label{Debye}
\left.\left.-2\zeta\left[\cos(g_m\xi_1)+\cos\left(\frac{g_m}{2}
(\xi_1+\sqrt{3}\xi_2)\right)
+\cos\left(\frac{g_m}{2}
(\xi_1-\sqrt{3}\xi_2)\right)\right]\right]\right\}.
\end{equation}
In particular, the Debye masses of the two independent fields 
($\xi_1$ and $\xi_2$ for our choice), 
following from Eq.~(\ref{Debye}), turn out 
to be equal to each other and read $m=g_m\sqrt{3\zeta}$.

For bookkeeping purposes, 
note that Eq.~(\ref{3}) can also be represented in the form 
of the $SU(3)$-inspired 
Toda type theory~\cite{toda}. 
This can be done by introducing  
the vectors 
$\vec\eta_1=\left(-\frac12,\frac{1}{2\sqrt{3}}\right)$,
$\vec\eta_2=\left(\frac12,\frac{1}{2\sqrt{3}}\right)$,
$\vec\eta_3=\left(0,-\frac{1}{\sqrt{3}}\right)$, which are  
just the weights of the representation $[3]$ of ${\,}^{*}SU(3)$.
Then, in terms of the 
fields $\bar\chi_\alpha\equiv\vec \eta_\alpha \vec\chi$, 
the partition function~(\ref{3}) takes the form ({\it cf.} Ref.~\cite{2})

\begin{equation}
\label{4}
{\cal Z}=\int\left(\prod\limits_{\alpha=1}^{3}{\cal D}\bar\chi_\alpha
\right)\delta\left(\sum\limits_{\alpha=1}^{3}
\bar\chi_\alpha\right)
\exp\left\{-\int d^3x
\left[\left(\nabla\bar\chi_\alpha\right)^2-
2\zeta\sum\limits_{{\alpha,\alpha'=1}\atop{\alpha>\alpha'}}^{3}
\cos\left(g_m\left(\bar\chi_\alpha-\bar\chi_{\alpha'}\right)
\right)\right]\right\}.
\end{equation}
 
However, for constructing the string representation 
of the Wilson loop in the theory~(\ref{aux}) (or~(\ref{3})), 
the representation~(\ref{4}) 
will not be necessary. To construct this representation, it is 
first convenient to derive an expression for the partition 
function~(\ref{aux}) in terms of the integral over monopole 
densities. This procedure is analogous to the one employed 
in Ref.~\cite{eur} for the case of compact QED. Firstly, 
let us multiply Eq.~(\ref{aux}) by the following unity: 
$1=\int {\cal D}\vec\rho{\,}\delta(\vec\rho(\vec x)-\vec\rho_{\rm gas}
(\vec x))$. After that, this equation reads

$${\cal Z}=\int {\cal D}\vec\lambda {\cal D}\vec\rho
\left\{ 
1+\sum\limits_{N=1}^{\infty}
\frac{\zeta^N}{N!}\left(\prod\limits_{a=1}^{N}
\int d^3z_a\sum\limits_{\alpha_a=\pm 1,\pm 2,\pm 3}^{}
\right)\times\right.$$

\begin{equation}
\label{5}
\left.
\times\exp\left[-\frac{2\pi}{g^2}\int d^3xd^3y\vec\rho(\vec x)
\frac{1}{|\vec x-\vec y{\,}|}\vec\rho(\vec y)\right]\right\}
\exp\left[-ig_m\int d^3x
\vec\lambda(\vec\rho-\vec\rho_{\rm gas})\right],
\end{equation}
where $\vec\lambda$ stands for the Lagrange multiplier.
Next, upon the normalization of 
the measure ${\cal D}\vec\lambda$ by the 
condition~\footnote{Clearly, this condition can be written as 
$\int {\cal D}
\vec\lambda\exp\left[-\frac12\int d^3x\left(
\nabla\vec\lambda\right)^2\right]=1$.}

$$
\int {\cal D}\vec\lambda {\cal D}\vec\rho\exp\left[
-\frac{2\pi}{g^2}\int d^3xd^3y\vec\rho(\vec x)
\frac{1}{|\vec x-\vec y{\,}|}\vec\rho(\vec y)-ig_m\int d^3x
\vec\lambda\vec\rho\right]=1,$$
Eq.~(\ref{5}) can be written as follows:

$${\cal Z}=\int 
{\cal D}\vec\lambda {\cal D}\vec\rho
\exp\left[-\frac{2\pi}{g^2}\int d^3xd^3y\vec\rho(\vec x)
\frac{1}{|\vec x-\vec y{\,}|}\vec\rho(\vec y)
-ig_m\int d^3x\vec\lambda\vec\rho\right]\times$$

$$\times\left\{ 
1+\sum\limits_{N=1}^{\infty}
\frac{\zeta^N}{N!}\left(\prod\limits_{a=1}^{N}
\int d^3z_a\sum\limits_{\alpha_a=\pm 1,\pm 2,\pm 3}^{}
\right)
\exp\left(ig_m\int d^3x\vec\lambda\vec\rho_{\rm gas}\right)
\right\}=$$

\begin{equation}
\label{neweq}
=\int {\cal D}\vec\rho{\,} {\cal D}\vec\lambda
\exp\left\{-\frac{2\pi}{g^2}\int d^3xd^3y\vec\rho(\vec x)
\frac{1}{|\vec x-\vec y{\,}|}\vec\rho(\vec y)
+\int d^3x\left[
2\zeta\sum\limits_{\alpha=1}^{3}\cos\left(
g_m\vec q_\alpha\vec\lambda\right)
-ig_m\vec\lambda\vec\rho
\right]\right\}.
\end{equation}
The integration over $\vec\lambda$ reduces now 
to the problem of finding a solution to the following 
saddle-point equation:

\begin{equation}
\label{sp}
\sum\limits_{\alpha=1}^{3}\vec q_\alpha\sin\left(g_m\vec q_\alpha
\vec\lambda\right)=-\frac{i\vec\rho}{2\zeta}.
\end{equation}
This equation can be solved {\it w.r.t.} $\vec q_\alpha\vec\lambda$
by noting that an arbitrary vector $\vec\rho\equiv
(\rho^1,\rho^2)$ can always be represented as 
$\sum\limits_{\alpha=1}^{3}\vec q_\alpha\rho_\alpha$, where 
$\rho_1=\frac{1}{\sqrt{3}}\left(\frac{1}{\sqrt{3}}\rho^1+\rho^2\right)$,
$\rho_2=-\frac23\rho^1$, 
$\rho_3=\frac{1}{\sqrt{3}}\left(\frac{1}{\sqrt{3}}\rho^1-\rho^2\right)$.
Then, inserting the so-obtained expression for $\vec q_\alpha\vec\lambda$ 
back into the action standing in the 
argument of the exponent  
on the R.H.S. of Eq.~(\ref{neweq}), we eventually arrive at  
the following 
representation for the partition function~(\ref{aux}) in terms 
of the monopole densities:

\begin{equation}
\label{multi}
{\cal Z}=\int {\cal D}\vec\rho\exp\Biggl\{-\Biggl[
\frac{2\pi}{g^2}\int d^3xd^3y\vec\rho(\vec x)\frac{1}{|\vec x-
\vec y{\,}|}\vec\rho(\vec y)
+V[\vec\rho{\,}]\Biggr]\Biggr\}.
\end{equation}
Here, the effective multivalued monopole potential $V[\vec\rho{\,}]$  
has the form

\begin{equation}
\label{pot}
V[\vec\rho{\,}]=\sum\limits_{n=-\infty}^{+\infty}
\sum\limits_{\alpha=1}^{3}\int d^3x
\left\{\rho_\alpha\left[\ln\left(\frac{\rho_\alpha}{2\zeta}+
\sqrt{1+\left(\frac{\rho_\alpha}{2\zeta}\right)^2}\right)+2\pi in\right]-
2\zeta\sqrt{1+\left(\frac{\rho_\alpha}{2\zeta}\right)^2}\right\}.
\end{equation}

We are now in a position to discuss the string representation of the 
Wilson loop. The contribution of diagonal gluons $\vec A_\mu\equiv
(A_\mu^3, A_\mu^8)$ into this 
quantity, which we are interested with, reads~\footnote{Here and in the 
next equation,  
$\vec\lambda$ again denotes the vector $(\lambda_3, \lambda_8)$.} 

\begin{equation}
\label{ini}
\left<W(C)\right>\equiv\frac13\left<{\rm tr}{\,}P{\,}\exp\left(
\frac{i}{2}\oint\limits_{C}^{}dx_\mu\vec A_\mu\vec\lambda\right)\right>.
\end{equation}
Clearly, since both $\lambda_3$ and $\lambda_8$ are diagonal,
the path ordering can be omitted (which becomes obvious from 
the definition of the path-ordering prescription). Owing to the 
Stokes theorem, one then obtains for the desired monopole contribution 
to the Wilson loop the following expression 
({\it cf.} Ref.~\cite{eur} for the case of compact QED):

$$
\left<W(C)\right>_{\rm m}=\frac13{\,}\left<{\rm tr}{\,}\exp\left(
\frac{i}{2}\int d^3x\vec\rho_{\rm gas}\vec\lambda\eta
\right)\right>.
$$
Here, 

\begin{equation}
\label{arbsolid}
\eta\left[\vec x,\Sigma\right]\equiv\frac12\varepsilon_{\mu\nu\lambda}
\frac{\partial}{\partial x_\mu}\int\limits_{\Sigma}^{}
d\sigma_{\nu\lambda}(\vec y)\frac{1}{|\vec x-\vec y{\,}|}
\end{equation}
denotes the solid angle, 
under which a certain surface $\Sigma$, bounded by the 
contour $C$, shows up to an observer located at the point $\vec x$.
(Clearly, $\eta$ is a function of the point $\vec x$ and a functional
of the surface $\Sigma$, {\it i.e.} of the vector $\vec y(\xi^1, 
\xi^2)$, which parametrizes this surface.)
Owing to the explicit form of the matrices $\lambda_3$ and 
$\lambda_8$, one finally obtains ({\it cf.} Ref.~\cite{2}) 

\begin{equation}
\label{newwil}
\left<W(C)\right>_{\rm m}=\frac13\left<\sum\limits_{\alpha=1}^{3}
\exp\left(i\int d^3x\vec\rho_{\rm gas}\vec\eta_\alpha\eta\right)
\right>.
\end{equation}

Noting now that the average in 
Eq.~(\ref{newwil}) is taken {\it w.r.t.} the monopole 
partition function~({\ref{aux}), 
we can apply to this equation 
the same procedure, which led to  
the representation~(\ref{multi})-(\ref{pot}). In this way,  
we conclude that the monopole contribution to the Wilson loop~(\ref{ini})
is given by the following expression:

\begin{equation}
\label{fin}
\left<W(C)\right>_{\rm m}=\frac{1}{3{\cal Z}}\sum\limits_{\alpha=1}^{3}
\int {\cal D}\vec\rho
\exp\Biggl\{-\Biggl[
\frac{2\pi}{g^2}\int d^3xd^3y\vec\rho(\vec x)\frac{1}{|\vec x-
\vec y{\,}|}\vec\rho(\vec y)
+V[\vec\rho{\,}]-
i\int d^3x\vec\rho\vec\eta_\alpha\eta\Biggr]\Biggr\}.
\end{equation}
Similarly to compact QED~\cite{confstr, eur},
a seeming $\Sigma$-dependence of the R.H.S. of this equation,
brought about by the solid angle, actually disappears due 
to the summation over all the complex-valued branches of the 
effective potential~(\ref{pot}) at every point $\vec x$.
This observation is the essence of the string 
representation of the Wilson loop in the monopole gas both 
in compact QED and in the $SU(3)$-case under study. 
The $SU(3)$-analogue of the so-called confining string 
theory, proposed for the case of compact QED 
in Ref.~\cite{confstr}, can be obtained by 
the following change of variables in the functional integral
standing on the R.H.S. of Eq.~(\ref{fin}): $\vec\rho\to\vec F_{\mu\nu}$.
Here, the monopole field strength tensor 
$\vec F_{\mu\nu}(\vec x)=-\varepsilon_{\mu\nu\lambda}
\partial_\lambda^x\int d^3y\frac{\vec\rho(\vec y)}{|\vec x-
\vec y{\,}|}$
obeys Bianchi identities,
modified by monopoles, 
$\frac12\varepsilon_{\mu\nu\lambda} 
\partial_\mu\vec F_{\nu\lambda}=4\pi\vec\rho$.
Such a substitution 
is obvious, and we will not discuss it here, referring the 
reader to Ref.~\cite{eur} for a detailed comparison 
of our approach with the theory of confining strings in the 
case of compact QED.
Note only that such a reformulation of the functional integral
allows one to account automatically also for the noncompact 
part of the $\vec A_\mu$-fields. Clearly, 
that is because $\vec F_{\mu\nu}$
is defined up to an addendum
$\partial_\mu\vec A_\nu-\partial_\nu\vec A_\mu$ with 
single-valued $\vec A_\mu$'s. 

The obtained string representation~(\ref{fin}) can now
be applied to the evaluation of the Wilson loop in the 
approximation when the monopole gas is dilute, {\it i.e.} 
$|\vec\rho{\,}|\ll\zeta$. In this way, we
can restrict ourselves to the real branch of the 
monopole potential~(\ref{pot}), provided that in Eq.~(\ref{arbsolid})
the replacement 
$\Sigma\to\Sigma_{\rm min.}$, with   
$\Sigma_{\rm min.}$ standing the surface of the minimal area
for a given contour $C$, has been performed. Then, the 
Wilson loop in the dilute monopole gas reads

$$\left<W(C)\right>_{\rm m}=\frac{1}{3
\left<W(0)\right>_{\rm m}}
\sum\limits_{\alpha=1}^{3}\int {\cal D}\vec\rho
\exp\Biggl\{-\Biggl[
\frac{2\pi}{g^2}\int d^3xd^3y\vec\rho(\vec x)\frac{1}{|\vec x-
\vec y{\,}|}\vec\rho(\vec y)+$$

\begin{equation}
\label{dil}
+\int d^3x\left(-6\zeta
+\frac{1}{6\zeta}\vec\rho{\,}^2-
i\vec\rho\vec\eta_\alpha\eta(\vec x)
\right)\Biggr]\Biggr\}=\exp\left\{-\frac{\zeta}{8\pi}\int d^3xd^3y
\frac{{\rm e}^{-m|\vec x-\vec y{\,}|}}{|\vec x-\vec y{\,}|}
\partial_\mu^x
\eta(\vec x)
\partial_\mu^y\eta(\vec y)
\right\}.
\end{equation}
Here, we have for brevity denoted $\eta(\vec x)\equiv
\eta\left[\vec x, \Sigma_{\rm min.}\right]$ and 
used the fact that for every ``$\alpha$'', 
$\vec\eta_\alpha{\,}^2=\frac13$. It is further worth employing 
the following formula for the derivative of the solid angle
(see {\it e.g.}~\cite{eur}):

$$\partial_\mu^x\eta\left[\vec x, \Sigma_{\rm min.}\right]=
\varepsilon_{\mu\nu\lambda}\left[\partial_\nu^x\oint\limits_{C}^{}
dy_\lambda\frac{1}{|\vec x-\vec y{\,}|}-2\pi\int
\limits_{\Sigma_{\rm min.}}^{}d\sigma_{\nu\lambda}(\vec y)
\delta(\vec x-\vec y)\right]$$
(which is actually valid for an arbitrary surface $\Sigma$, bounded by $C$),
after which the derivation of $\left<W(C)\right>_{\rm m}$ becomes
straightforward. Combining the so-obtained result with
the contribution to the 
Wilson loop, stemming from the noncompact (``photon'') part of the 
$\vec A_\mu$-fields, which according to Eq.~(\ref{ini}) has the form 

$$\left<W(C)\right>_{\rm ph}=\exp\left(-\frac{g^2}{24\pi}
\oint\limits_{C}^{}dx_\mu\oint\limits_{C}^{}dy_\mu\frac{1}{|\vec x-
\vec y{\,}|}\right),$$  
we finally obtain the following result for the full Wilson loop:

$$\left<W(C)\right>=\left<W(C)\right>_{\rm m}\left<W(C)\right>_{\rm ph}=$$

\begin{equation}
\label{wfull}
=\exp\left\{-\left[
\pi\zeta\int
\limits_{\Sigma_{\rm min.}}^{}d\sigma_{\mu\nu}(\vec x)
\int\limits_{\Sigma_{\rm min.}}^{}d\sigma_{\mu\nu}(\vec y)
\frac{{\rm e}^{-m|\vec x-\vec y{\,}|}}{|\vec x-\vec y{\,}|}
+\frac{g^2}{24\pi}\oint\limits_{C}^{}dx_\mu\oint\limits_{C}^{}dy_\mu
\frac{{\rm e}^{-m|\vec x-\vec y{\,}|}}{|\vec x-\vec y{\,}|}\right]\right\}.
\end{equation}
By virtue of the results of Ref.~\cite{expan}, it is easy to 
get the string tension of the Nambu-Goto term, which is the 
leading term in the gradient (or $1/m$-) expansion of the nonlocal string
effective action standing as the first argument of the exponent on the R.H.S.
of Eq.~(\ref{wfull})~\footnote{
Note that at the surface of the minimal 
area, the next-to-leading term of this expansion (the so-called 
rigidity term) vanishes~\cite{shep}.}. The string tension 
reads $\pi g\sqrt{\frac{\zeta}{3}}$
and is therefore nonanalytic in $g$ (owing to the nonanalyticity 
in the $g$-dependence of the fugacity), similarly to what happens in the 
real QCD. However, it remains unclear within the model under study
how to derive the fugacity itself from the QCD Lagrangian. 
Owing to this, the 
approach to the problem of 
string representation of QCD, elaborated in 
the present Letter, does not answer the question: what is the proportionality
coefficient between the string tension in QCD and $\Lambda_{\rm QCD}^2$?
This question, the answer to which is very important for understanding 
the connection between the perturbative and nonperturbative 
phenomena in QCD, will be addressed in future publications.

\acknowledgments
The author is indebted to Profs. A. Di Giacomo and Yu.A. Simonov
for useful discussions. 
He is also greatful to Prof. A. Di Giacomo and  
the whole staff of the Quantum Field Theory Division
of the University of Pisa for cordial hospitality and INFN for  
financial support.

\end{document}